\documentclass[12pt]{article}
\usepackage{epsfig}
\begin{document}
\title{A study of trapped Bose gas beyond mean-field theory}
\author{Arup Banerjee and Manoranjan P. Singh\\
Laser Physics Division, Centre for Advanced Technology\\
Indore 452013, India}

\maketitle
\begin{abstract}
We study the ground-state properties of dilute Bose gas confined to both isotropic and anisotropic traps to assess the accuracy of Gross-Pitaevskii (GP) theory. To go beyond 
GP approximation we use Huang-Yang theory of interatomic interaction energy
for hard-sphere Bose gas and use a variational method to solve the resulting modified GP
equation. We also make an analytic estimate of the corrections due to the higher-order
terms in the interatomic interaction energy. We find that corrections are of 
the order of $1\%$. However, there is a qualitative change in the density profile due to
presence of logarthimic term in the interaction energy for large $N$ (number of atoms).
\end{abstract}

\section{Introduction}
The discovery of Bose-Einstein condensation (BEC) in magnetically trapped alkali atoms
\cite{anderson,bradley,davis} and spin-polarized hydrogen atom \cite{fried} has resulted 
in large body of theoretical and experimental 
research \cite{dalfovo} on the ground-state as well as excited state properties of trapped atomic gases. For the case of dilute atomic gas, which is the case for most of the experimental situation, the mean-field theory of GP \cite{gp} has been quite successful in explaining both static (ground-state) and dynamic (collective excitations) properties of the condensates. The interacting atomic gas is considered to be dilute when the average distance between the atoms is much larger than the range of interaction potential between the atoms \cite{huang} (mathematically this is expressed by 
the condition $n|a|^{3} << 1$, where $n$ atomic density and $a$ is the
s wave scattering length of interatomic potential). For this case the interaction between the atoms
is well described by two-body collisions characterized by a single parameter obtainable
from the potential namely, the s wave scattering length $a$.

In some recent experiments \cite{fried,stamper2,stamper1} BEC is achieved with number of atoms $N$
being quite high. It is possible to produce condensates of $\approx 10^7$  $^{87}$Rb 
atoms, $\approx 5\times 10^7$ Na atoms and $\approx 10^9$ H atoms. The maximum density 
in the optical traps has been around 10$^{15}$ cm$^{-3}$. Thus, with the increasing 
atomic density in the condensate it becomes necessary to test the validity of mean-field 
GP theory. The main aim of this paper is to investigate systematically the accuracy of 
the GP theory in the high density limit. We go beyond mean-field theory and evaluate the 
corrections using a variational formalism. We also derive an analytic estimate for the 
corrections in the frame work of Thomas-Fermi(TF) approximation. For this purpose we use Huang-Yang
\cite{hy} theory for the interatomic interaction energy $\epsilon$. In this theory, 
$\epsilon$ is expanded in power of gas parameter $n|a|^{3}$ by assuming the atomic gas to be
a uniform hard-sphere Bose gas. It is important to note here that the interatomic potential in Huang-Yang theory is replaced by a pseudopotential. The expression for $\epsilon$ is
\cite{lee1,lee2}
\begin{equation}
\epsilon(n) = \frac{2\pi\hbar^{2}an}{m}\left [1 + \frac{128}{15\sqrt{\pi}}\left (na^{3}
\right )^{\frac{1}{2}} + 8\left (\frac{4\pi}{3} - \sqrt{3}\right )(na^{3})\ln\left (na^{3}\right )
+ {\cal O}\left(na^{3}\right)\right ]
\label{1}
\end{equation}
In GP theory only the first term in  the above expansion is considered as higher order terms are negligible for dilute gas. However with increasing density higher-order
terms become appreciable. It is, therefore, of interest to study the effects these higher 
order terms on the ground-state properties of BEC.

The effect of higher-order terms in the interatomic potential on the physical properties 
of BEC has recently been reported in the literature \cite{nunes,polls}.
Nunes \cite{nunes} has used density functional theory to derive 
the modified GP (MGP) equation. By solving the MGP equation for a weak isotropic trap 
in the Thomas-Fermi (TF) approximaion, he finds that 
the maximum change in the ground state density due to the higher order terms in 
Eq.(\ref{1})is only a few percent for the value of $N$ 
as large as $10^9$. On the other hand, the change in the ground state energy per particle 
has been reported to be sensitive to $N$ - for $N \le 10^7$ both GP and MGP results are 
very close whereas, they differ by 33$\%$ for $N=10^9$. This result is not in
conformity with the 
analytic estimate \cite{dalfovo} of the correction introduced in the ground state energy
by the higher order terms in Eq.(\ref{1}). According to the analytic estimate, the first 
higher order term in Eq.(\ref{1}) causes a change of only a few percent in the ground state 
energy even for very large value of $N$. As discussed below, the second higher order term 
in Eq.(\ref{1}) being logarithmic gives negative correction. Therefore, this is expected 
to partially cancel the correction due to the first higher-order term in the interatomic
interaction energy. As a result of this, the net 
correction remain close to a few percent when $N$ is varied from $10^7$ to $10^9$. 
Fabrocini and Polls 
 \cite{polls} have studied the effect of interatomic correlations within the local density 
approximation and the correlated basis function (CBF) theory. They have solved the resulting 
MGP equation and the lowest order correlated Hartree (CH$_{LO}$) equation for an isotropic trap using 
the steepest descent method. Both of these methods give corrections which are close to those 
predicted by the analytical estimates. However, the CH$_{LO}$ equation gives lower value of the ground state 
energy than the MGP equation. The difference between the results obtained by these two
methods shows up more at higher values of $N$. It is not clear if this is due to any 
numerical problem or due to some physical reason. We discuss more about these results in 
section 4.

 In this paper we employ variational approach \cite{byam,fetter,mp} to solve the MGP 
equation by using the variational ansatz for the density recently proposed in 
Ref.\cite{mp}. We also perform calculations for more realistic case of axially symmetric 
trap ($\omega_{x} = \omega_{y} = \omega_{\bot} \neq \omega_{z}$) potential. In order to 
assess the accuracy of our variational results we generalize the virial relations 
\cite{dalfovo} for the interatomic interaction energy given by Eq.(\ref{1}). 

Recently, several authors \cite{dalfovo,timmermans,braaten} have estimated the corrections
introduced by the first higher-order term in the Eq.(\ref{1}) to assess the accuracy of 
GP theory. These authors obtained analytical expressions for the corrections over 
mean-field results within the Thomas-Fermi (TF) approximation valid in the large $N$ limit.
Here we extend these results by including second higher-order term of Eq.(\ref{1}). 
Beside, providing the estimates for corrections these results also serve as a check for 
the accuracy of our numbers obtained by the variational calculation.

It is necessary to emphasize here that our previous and the present study clearly bring
out the advantages of variational approach. This approach requires numerically less effort
than the direct numerical integration method and can handle system of arbitrary number of 
particle. At the same time, it provides a very accurate description of the ground state 
properties of the system.

Rest of the paper is organized as follows: In section 2 we describe the variational 
approach and also derive an expression for the generalized virial relation. Section 3 
contains analytical estimates for corrections in the density profile, the total energy and 
the chemical potential. Results are discussed in section 4 and paper is concluded in 
section 5.

\section{Variational method}
The energy functional for a system of N-bosons, each of mass $m$, confined in a trap 
potential $V_{t}({\bf r})$ is given by
\begin{equation}
E[\Psi] = \int d{\bf r}\left [\frac{\hbar^{2}}{2m}|{\bf\nabla}\Psi({\bf r})|^{2} + 
V_{t}({\bf r})|\Psi({\bf r})|^{2} + \epsilon(n)|\Psi({\bf r})|^{2} \right ],
\label{2}
\end{equation}
where $\Psi({\bf r})$ is the corresponding wave function and gives density via the relation
$n({\bf r}) = |\Psi({\bf r})|^{2}$. For $\epsilon(n)$ we use the local-density 
approximation (LDA) expression as given in Eq.(\ref{1}). The energy functional associated 
with GP theory can be obtained from Eq.({\ref2}) by substituting only the first term of 
Eq.(\ref{1}) for $\epsilon(n)$ in Eq.(\ref{2}).
In this study we go beyond GP theory by including the second and the third term of 
Eq.(\ref{1}) to study the ground-state properties of BEC. 

For our purpose we consider an axially symmetric trap characterized by two angular frequencies
$\omega_{\bot}^{0}$ and $\omega_{z}^{0}$ and the corresponding potential is given by
\begin{equation}
V_{t}({\bf r}) = \frac{m{\omega_{\bot}^{0}}^{2}}{2}\left (x^{2} + y^{2} + \lambda_{0}^{2}z^{2}
\right ),
\label{3}
\end{equation}
where $\lambda_{0} = \omega_{\bot}^{0}/\omega_{z}^{0}$ is the anisotropy parameter of 
the trapping potential.

As in the GP approach, from Eqs.(\ref{1}) and (\ref{2}) the MGP equation can be derived by making
the energy functional $E[\Psi]$ stationary with respect to the wave function 
$\Psi({\bf r})$ such that the wave function remains normalized to the total number of 
particles $N$ given by the condition
\begin{equation}
\int|\Psi({\bf r})|^{2}d{\bf r} = N.
\label{norm}
\end{equation}
 Such a variation leads to:
\begin{eqnarray}
&  &\left [ - \nabla_{1}^2 
+  \left( x_{1}^2 + y_{1}^2 + \lambda^{2}_{0} z_{1}^2 \right) 
+8\pi\overline{a}N |\psi_{1}({\bf r}_{1})|^2 + \frac{256}{3}\left(\pi\overline{a}^5N^3\right )
^{\frac{1}{2}}|\psi_{1}({\bf r}_{1})|^{3}\right. \nonumber \\
 &  & + \left.  8\pi\overline{a}^{4}N^{2}\left (\frac{4\pi}{3} - \sqrt{3}\right )
|\psi_{1}({\bf r}_{1})|^{4}
\left\{6\ln\left (N\overline{a}^{3}|\psi_{1}({\bf r}_{1})|^{2}\right ) + 2\right\}\right ]
\psi_{1}({\bf r}_{1}) \nonumber \\
&  & = 2 \mu_{1} \psi_{1}({\bf r}_{1})
\label{4}
\end{eqnarray}
where, $\mu_{1}$ is the chemical potential in the units of $\hbar\omega_{\bot}^{0}$ 
corresponding to the normalization constraint given by Eq.(\ref{norm}). In  this equation we have used the length scale rescaled
variables ${\bf r}_{1} = {\bf r}/a_{ho}$,$\overline{a}=a/a_{ho}$,
${\bf \nabla_{1}}=a_{ho}{\bf \nabla}$, 
$E_{1} = E/\hbar\omega_{\bot}^{0}$, with
$a_{ho} = \left (\hbar/m\omega_{\bot}^{o}\right )^{\frac{1}{2}}$ and $\Psi({\bf r}_{1})$
is normalized to unity.

In Ref.\cite{polls} and \cite{nunes} above equation (Eq.\ref{4}) has been solved 
numerically for isotropic trap $(\lambda_{0} = 1)$. In this paper we take recourse to 
variational approach to solve the Eq.(\ref{4}). The main advantage of this method is that 
with a suitable choice for the variational form of the wave function one can get quite 
accurate results with less computational effort. In addition, this approach may provide 
physical insight which generally get obscured by the complicated computational procedures. 
In this paper we use the variational form given by \cite{mp}
\begin{equation}
\psi_{1}({\bf r}_{1}) = \sqrt{\frac{p}{2 \pi \Gamma(\frac{3}{2 p})}}
\lambda^{\frac{1}{4}} \left(\frac{\omega_{\bot}}{\omega_{\bot}^0}\right)
^{\frac{3}{4}} e^{-\frac{1}{2} 
 \left(\frac{\omega_{\bot}}{\omega_{\bot}^0}\right)^p
\left( r_{1 \bot}^2 +\lambda z_1^2\right)^p}
\label{5}
\end{equation}
where $\lambda$, $\omega_{\bot}$ and $p$ are the variational parameters which are obtained
by minimizing $E_{1}$ with respect to this parameters. This variational form has been shown
to describe quite accurately the ground state of  the dilute Bose gas confined in a trap 
for a wide range
of particle numbers. When number is small it tends correctly towards Gaussian and in the 
opposite limit it resembles with the TF wave function. In addition to this in the 
intermediate region it combines the feature of both in an effective way. This variational 
form is also well suited for negative scattering length and is easily generalized for the 
vortex states. Further, using Eq.(\ref{5}) the
physical observables can be expressed analytically in terms of these three variational
parameters. For example, energy functional $E_{1}$ can be written as:
\begin{equation}
\frac{E_{1}}{N}=E_{kin}+E_{ho}+E_{int}^1+E_{int}^{2}+E_{int}^{3}.
\label{energy}
\end{equation}
Here $E_{kin}$, $E_{ho}$ denote average kinetic and harmonic trapping potential energies, 
respectively. Remaining three terms $E_{int}^{1}, E_{int}^{2}$ and $E_{int}^{3}$ represent 
interatomic interaction energies corresponding to the first, second and the third terms in
the energy expansion in Eq.(\ref{1}).
\begin{equation}
E_{kin}  =  \frac{1}{12}\frac{\omega_{\bot}}{\omega^0_{\bot}}
\left(1+\frac{\lambda}{2}\right)
\frac{\Gamma\left( \frac{1}{2p}\right)}{\Gamma\left(\frac{3}{2p}\right)}
(1+2p)
\end{equation}
\begin{equation}
E_{ho}=\frac{1}{3}\frac{\omega^0_{\bot}}{\omega_{\bot}}
\left(1+\frac{\lambda_0^2}{2\lambda}\right)
\frac{\Gamma\left( \frac{5}{2p}\right)}{\Gamma\left(\frac{3}{2p}\right)}
\end{equation}
\begin{equation}
E_{int}^1= N\frac{a}{a_{\bot}}
\left(\frac{\omega_{\bot}}{\omega^0_{\bot}}\right)^{\frac{3}{2}}
\sqrt{\lambda}
\frac{p)}{\Gamma\left(\frac{3}{2p}\right)}
\left(\frac{1}{2}\right)^{\frac{3}{2p}}
\end{equation}
\begin{equation}
E_{int}^2= \frac{128}{15\sqrt{2\pi}}\frac{a^{\frac{5}{2}}N^{\frac{3}{2}}p^{\frac{3}{2}}
\lambda^{\frac{3}{4}}\left (\frac{\omega_{\bot}}{\omega_{\bot}^{0}}\right )}
{\left (\Gamma\left (\frac{3}{2p}\right )\right )^{\frac{3}{2}}}\left (\frac{2}{5}\right )
^{\frac{3}{2p}}
\end{equation}
\begin{eqnarray}
E_{int}^3&= & 4\overline{a}^{4}N^{2}\left (\frac{4\pi}{3} - \sqrt{3}\right )
\frac{p^{2}\lambda
\left (\frac{\omega_{\bot}}{\omega_{\bot}^{0}}\right )^{3}}{\pi
\left (\Gamma\left (\frac{3}{2p}\right )\right )^{2}3^{\frac{3}{2p}}} \times \\ \nonumber
&&\left[ \ln\left [N\overline{a}^{3}\left (\sqrt{\frac{p}{2\pi\Gamma\left (\frac{3}{2p}
\right)}}\lambda^{\frac{1}{4}}\left (\frac{\omega_{\bot}}{\omega_{\bot}^{0}}\right )
^{\frac{3}{4}}\right )^{2}\right ]
 - \left (\frac{\omega_{\bot}}{\omega_{\bot}^{0}}\right )^
{\frac{p}{2}}\frac{1}{2\sqrt{3}}\Gamma\left (\frac{3}{2p} + 1
\right )\right]
\end{eqnarray}
For particular value
$N$ the parameter $\omega_{\bot}$, $\lambda$ and $p$ are determined by minimizing the 
energy given above (Eq.(\ref{energy})).

The virial relation among different energy components provides a way of checking the 
correctness of variational or numerical solutions. For example, within GP theory different
energy components satisfy the virial relation \cite{dalfovo}
\begin{equation}
2E_{kin} - 2E_{ho} + 3E_{int}^{1} = 0
\label{virialgp}
\end{equation}
For our purpose we generalize the above relation (Eq.(\ref{virialgp})) valid for the MGP 
theory. By using variational nature of energy and the scaling 
transformation $\Psi({\bf r}_{\lambda})\rightarrow \sqrt{\lambda}\Psi({\bf r})$ we arrive 
at following expression for the virial relation corresponding to the MGP energy 
functional
\begin{equation}
2E_{kin} - 2E_{ho} + 3E_{int}^{1} + \frac{9}{2}E_{int}^{2} + 2E_{int}^{3} + \int |\Psi_{1}
({\bf r}_{1})|^{6}d{\bf r}_{1} = 0.
\label{6}
\end{equation}
 For each term of this expansion we have checked the 
satisfaction of virial relation corresponding to our variational solutions. The virial 
relation is satisfied up to 5-th decimal place in our calculation.

Before presenting the results, we make an estimate of the corrections in the density, 
total energy and the chemical potential as a result of inclusion of two new terms in the 
interatomic energy. For this we follow the approach of Dalfovo et al.\cite{dalfovo}, 
which is valid for large $N$.
 
\section{Analytic Estimate}
We use local density approximation for the chemical potential which is justified in the 
 thermodynamic limit $N \rightarrow \infty$ where the profile of the density distribution 
is very smooth. In this approximation the chemical potential is given by:
\begin{equation}
\mu=\mu_{local}[n({\bf r})] + V_t({\bf r})
\label{mu}
\end{equation}
Given the expression for $\mu_{local}(n)$ for the homogeneous Bose-gas, the density 
$n({\bf r})$ can be calculated by solving Eq.(\ref{mu}), along with the normalization condition which fixes 
the parameter $\mu$ on the left hand side of the equation. From Eq.(\ref{1}), it follows 
that
\begin{equation}
\mu_{local}(n) = gn\left [1 + \frac{32}{5\sqrt{\pi}}\left (na^{3}\right )^{\frac{1}{2}} + 4\left
 (\frac{4\pi}{3} - \sqrt{3}\right )\left (na^{3}\right )\left (3\ln \left (na^{3}\right ) 
\right ) + {\cal O}\left (na^{3}\right ) \right ],
\label{mul}
\end{equation}
where, 
\begin{equation}
g=\frac{4\pi \hbar^2 a}{m}.
\end{equation}
Considering only the first term in the expansion for $\mu_{local}$ we recover the mean-
field Thomas-Fermi result 
\begin{equation}
n({\bf r}) =  \frac{\left (\mu - V_{trap}({\bf r})\right )}
{g} = n_{TF}({\bf r}).
\label{7}
\end{equation}
Normalization condition implies that
\begin{equation}
\mu = \frac{\hbar\omega_{\bot}^{0}}{2}\left (15N\lambda_{0}\overline{a}\right )
^{\frac{2}{5}}=\mu_{TF}.
\label{9}
\end{equation}
Using the relation $\mu=\partial E/\partial N$ we get the corresponding expression for the 
ground sate energy as
\begin{equation}
E = \frac{5}{7}N\mu=E_{TF}.
\label{8}
\end{equation}
When the higher order terms in the expansion for $\mu_{local}$ are considered the standard 
procedure to solve Eq.(\ref{mu}) is by iteration \cite{dalfovo}. On following such procedure
 we get 
\begin{eqnarray}
n({\bf r})&= &\frac{\left (\mu - V_{t}({\bf r})\right )}
{g} -\frac{4 m^{3/2}}{3 \pi^2 \hbar^3}\left (\mu - V_{t}({\bf r})\right )^{3/2} \nonumber \\
& -&3\left
 (\frac{4\pi}{3} - \sqrt{3}\right )\frac{m^2 a}{4 \pi^2 \hbar^4} 
\left (\mu - V_{t}({\bf r})\right )^2 \ln\left(\frac{m a^2}{4 \pi \hbar^2}
 \left (\mu - V_{t}({\bf r})\right )\right)+{\cal O}(n a^3).
\label{dencorr}
\end{eqnarray}
Now imposing the normalization condition and neglecting the terms of order
${\cal O}(na^{3})$and higher, we obtain
\begin{equation}
\mu = \mu_{TF}\left [1 + \sqrt{\pi a^{3}n(0)} + \frac{96}{35}
\left (\frac{4\pi}{3} - \sqrt{3}\right )(n(0)a^{3})\ln\left (n(0)a^{3}\right )
\right ],
 \label{10}
\end{equation}
and consequently
\begin{equation}
E = \frac{5}{7}N\mu_{TF}\left [1 + \frac{7}{8}\sqrt{\pi a^{3}n(0)} + \frac{32}{15}
\left (\frac{4\pi}{3} - \sqrt{3}\right )(n(0)a^{3})\ln\left (n(0)a^{3}\right )\right ]
 \label{11}
\end{equation}
The parameter $a^{3}n(0)$ appearing in the above equations can be written as
\begin{equation}
a^{3}n(0) = \frac{15^{\frac{2}{5}}}{8\pi}\left (N^{\frac{1}{6}}\lambda_{0}^{\frac{1}{6}}
\overline{a}\right )^{\frac{12}{5}}
\label{12}
\end{equation}
Thus by using Eqs.(\ref{dencorr}),(\ref{10}) and (\ref{11}) we estimate the correction 
over the corresponding TF results.
For example a BEC with $N = 10^{7}$ particles and $\overline{a} = 4.33\times 10^{-3}$, the second terms
in Eq.(\ref{10}) and (\ref{11})
result in $2.2\%$ and $1.9\%$ corrections in the chemical potential and the total energy, 
respectively, over the corresponding TF values. Third terms, however, lead to negative 
corrections and these are $0.93\%$ in the chemical potential and $0.72\%$ in the total 
energy. Due to the cancellation the net corrections are approximately of order $1.2\%$ for 
both chemical potential and total energy over their corresponding TF values. Thus the 
third term which is logarithmic in nature compensates for the correction due to the second 
term in both chemical potential and energy. 

In the next section we present results of our variational calculation along with the 
numbers obtained from these expressions as a further check for our numerical results.

\section{Results and Discussion}

 In this section we first report calculations on $^{87}$Rb atoms confined in an isotropic 
$(\lambda_{0}=1)$ trap. The choice of isotropic trap is motivated only for the sake of 
comparison with the existing studies \cite{polls}. As pointed out earlier, our variational 
approach is capable of dealing with asymmetric traps with cylindrical symmetry or even 
fully symmetric traps. For the symmetric trap  we choose trap frequency $\omega_{0}/2\pi$ and 
scaled s wave scattering length $\overline{a}$ to be $77.78Hz$ and $4.33\times 10^{-3}$, 
respectively. In Table 1 we show chemical potentials, total energy and mean radius 
$<r^{2}>$ for particle numbers ranging from $10^{3}$ to $10^{9}$ for both GP and MGP cases.
For comparison we also show the corresponding MGP results of Ref.\cite{polls}in parenthesis.
\begin{table}
\caption{Results for the ground-state of $^{87}$Rb atoms  confined in an isotropic trap with
$\omega_{\bot}^{0}/2\pi = 77.78$ Hz. Chemical potentials and energies are in units of 
$\hbar\omega_{\bot}^{0}$ and length is in units of $a_{ho}$. Numbers in the brackets correspond
to the results of reference \cite{polls}}
\begin{center}
\begin{tabular}{|c|c|c|c|c|c|c|}\hline
N & \multicolumn{3}{c|}{GP} &
\multicolumn{3}{c|}{MGP}  \\ \cline{2-7}
& $\mu_{1}$ & E$_{1}$ & $<r>_{max}$ & $\mu_{1}$ & E$_{1}$ & $<r>_{max}$ \\
\hline

10$^{3}$ & 3.05 & 2.43 & 1.65 & 3.05 & 2.43 & 1.66 \\
         &      &      &       & (3.06) & (2.43) & (1.66)\\
10$^{4}$ & 6.87 & 5.05 & 2.44 & 6.91 & 5.07 & 2.45 \\
          &     &       &     & (6.92)  & (5.08) & (2.45)\\ 
10$^{5}$ & 16.90 & 12.13 & 3.81 & 17.02 & 12.21 & 3.83 \\
   &         &         &         & ( 17.07) & (12.25) &(3.84)\\
10$^{6}$ & 42.30 & 30.24 & 6.02 & 42.71& 30.51 &6.06 \\
        &       &        &      & (42.97) & (30.66) & (6.10)\\
10$^{7}$ & 106.18 & 75.86 & 9.54 & 107.53 & 76.75 & 9.61 \\ 
         &        &       &      & (108.75)  & (77.48)& (9.74)\\
10$^{8}$ & 266.70 & 190.50 & 15.12 & 270.42 & 193.13 & 15.23 \\
   &         &         &         & (275.89) & (196.45) & (15.45) \\
10$^{9}$ & 669.90 & 478.50 & 23.96 & 675.77 & 483.81 & 24.00 \\
   &         &         &         & - & - & - \\
\hline
\end{tabular}
\end{center}
\end{table}
It clearly shows that the difference
between GP and MGP results grow with the increase in particle number. For example, 
corresponding to $N = 10^{4}$ the GP numbers are lower than the MGP values by $0.4\%$. 
On the other hand for $N = 10^{8}$ the difference is of the order of $1.4\%$. This is 
consistent with our estimation of the corrections given by Eq.(\ref{10}) and (\ref{11}). 
Moreover, at this point we also
note that numbers obtained by us for total energy are systematically lower than the 
corresponding MGP results of Ref.\cite{polls}. The difference between our results and 
that of Ref.\cite{polls} increases with the number of particles. For example the difference is of 
the order $2\%$  at $N = 10^{7}$. A similar trend is observed for both chemical potential 
and mean radius. At this point it is worth mentioning that our method, being variational, 
provides an upper bound to the ground state energy. It is expected that the numerical value 
of the ground state energy obtained variationally will be typically higher than those 
obtained by solving 
the MGP equation directly by standard numerical procedures. It is therefore surprising that 
the results obtained by solving the MGP equation numerically in Ref. \cite{polls} gives 
higher value of the 
ground state energy than that obtained by us. On the other hand, numbers obtained by the
correlated wave function approach in Ref.\cite{polls} are very close to our results for
$N \leq 10^{6}$. For higher values of $N$, however, the trend is similar to the MGP results.

Next we compare results obtained by us with those of Ref.\cite{nunes}. To do this we consider 
BEC of $^{87}$Rb atoms in a trap with angular fequency of $20$Hz and $a = 7$nm. We do 
calculation for $N = 10^{9}$ as for this value of $N$ quite a drastic energy change is 
reported in Ref.\cite{nunes}. We find that for above parameters the energy per particle
corresponding
to the GP case is lower than the MGP value by $1.38\%$. This is also consistent with the 
analytical estimate given by Eq.(\ref{11}). The corresponding number reported in 
Ref.\cite{nunes} is $33\%$. We feel that this is a gross overestimate of the energy 
correction since the first higher-order term in the Eq.(\ref{11}) leads to the correction of around
$3\%$ over the TF value. But as mentioned before the second higher-order term in the 
Eq.(\ref{11}) being
logarthimic, results in a negative correction and for above parameters it is $1.6\%$ of
TF value. Consequently the net correction in energy per particle due to higher-order terms
is only $1.4\%$ over the TF number. We point out that this number compares well with the 
result obtained by variational calculation. 

In Table 2 we present the numbers for chemical potential and total energies obtained via 
Eq.(\ref{10}) and (\ref{11}), respectively for several values of N.  These numbers serve 
as an additional check for correctness of our numerical results
especially for large N. This is clearly the case as MGP numbers in Table 1 are quite close 
to the corresponding numbers in Table 2 for large N. Furthermore, comparison of Table 1
and II also shows that our variational results are closer to the analytical estimates than
those reported in Ref.\cite{polls}.

\begin{table}
\caption{Analytical estimates for the chemical potenial and the ground-state energy of 
$^{87}$Rb atoms confined in an isotropic trap with
$\omega_{\bot}^{0}/2\pi = 77.78$ Hz. Chemical potentials and energies are in units of 
$\hbar\omega_{\bot}^{0}$ and length is in units of $a_{ho}$. TF represents results within 
Thomas-Fermi approximation (Eqs.(\ref{9}) and (\ref{8})) and MTF corresponds to those obtained 
by modified TF expressions (Eqs.(\ref{10}) and (\ref{11}))}  
\begin{center}
\begin{tabular}{|c|c|c|c|c|}\hline
N & \multicolumn{2}{c|}{TF} &
\multicolumn{2}{c|}{MTF}  \\ \cline{2-5}
& $\mu_{1}$ & E$_{1}$  & $\mu_{1}$ & E$_{1}$  \\
\hline

10$^{3}$ & 2.66 & 1.90 & 2.66 & 1.90  \\
10$^{4}$ & 6.67 & 4.76 &  6.70 & 4.78  \\
10$^{5}$ & 16.75 & 11.96 & 16.87 & 12.04  \\
10$^{6}$ & 42.07 & 30.05 &  42.49 & 30.33 \\
10$^{7}$ & 105.68 & 75.49 & 107.05 & 76.41 \\ 
10$^{8}$ & 265.46 & 189.62 & 269.28 & 192.38  \\
10$^{9}$ & 666.81 & 476.29 & 673.19 & 482.42  \\
\hline
\end{tabular}
\end{center}
\end{table}
Next to study the density profile we plot the ground-state wave function along x-axis in 
Fig. 1 corresponding to both MGP (solid line) and GP (dotted line) 
cases for three different values of N: N $= 10^{7}, 10^{8}$ and $10^{9}$. It clearly shows 
that for all N the difference between the GP and the MGP wave functions arises at the region
where it reaches maximum, that is at the bottom of the potential well. Moreover, for 
N $=10^{7}$ (Fig. 1a) and $10^{8}$ (Fig. 1b) the value of wave function near origin 
corresponding to the GP case is higher than that of MGP case. The difference is of the order 
$1\%$ only. However, the situation becomes just opposite for the case of N $= 10^{9}$ 
(Fig. 1c). The reason for such change in
wave function profile is the contribution of logarithmic term in the interatomic energy
becoming appreciable at this N. As is already discussed that this term introduces
negative correction to the interatomic energy and potential.  Thus the logarithmic term 
gives rise to an attractive potential for the atoms in BEC at large N. Consequently, 
inclusion of this term in the interatomic energy (potential) brings atoms closer to the 
bottom of potential leading to the fact that more atoms are being found near origin than 
the GP case. Finally it is important to note here that the tail region of the wave function
is not affected by the higher-order terms in the interatomic potential.

After having discussed results for isotropic trap we now describe the results for 
anisotropic trap which is more relevant from experimental point of view. For this 
calculation we employ the numbers for the asymmetry parameter and the axial frequency 
corresponding to the experiment of Anderson et al. \cite{anderson}. Accordingly, 
$\lambda_{0} = \sqrt{8}$ and $\omega_{z}^{0}/2\pi = 220Hz$.
The value of s-wave scattering length is same as that of isotropic case considered earlier.
We present these results in Table 3. In this Table we compare the MGP results with the 
corresponding GP numbers. Similar to the isotropic case here also the virial relation 
Eq.(\ref{6}) is satisfied up to 5-th decimal place. The trend in both chemical potential 
and total energy with increase in the number of particles is similar to that of isotropic 
case. The difference between the MGP and GP numbers are of the same order as that of 
isotropic case. Moreover, in the axially symmetric case also we find that our variational 
numbers are quite close to the numbers obtained via Eq.(\ref{10}) and (\ref{11}) 
especially for $N > 10^{6}$. 
\begin{table}
\caption{Results for the ground-state of $^{87}$Rb atoms  confined in anisotropic trap 
with $\lambda_{0} = \sqrt{8}$ and
$\omega_{\bot}^{0}/2\pi = 77.78$ Hz.Chemical potentials and energies are in units of
$\hbar\omega_{\bot}^{0}$ and length is in units of $a_{ho}$.} 
\begin{center}
\begin{tabular}{|c|c|c|c|c|}\hline
N & \multicolumn{2}{c|}{GP} &
\multicolumn{2}{c|}{MGP}  \\ \cline{2-5}
& $\mu_{1}$ & E$_{1}$  & $\mu_{1}$ & E$_{1}$  \\
\hline
10$^{3}$ & 4.79 & 3.85 & 4.80 & 3.86  \\
10$^{4}$ & 10.53 & 7.78 &  10.58 & 7.82  \\
10$^{5}$ & 25.66 & 18.47 & 25.87 & 18.60  \\
10$^{6}$ & 64.13 & 45.86 &  64.84 & 46.33 \\
10$^{7}$ & 160.95 & 114.99 & 163.15 & 116.48 \\ 
10$^{8}$ & 404.24 & 288.75 & 409.72 & 292.66  \\
10$^{9}$ & 1015.38 & 725.27 & 1016.59 & 729.65  \\
\hline
\end{tabular}
\end{center}
\end{table}

Thus we conclude that the variational results obtained for anisotropic trap are also quite
accurate. Therefore, we demonstrate that by using variational approach described in this 
paper along with a judicious choice of ansatz for the ground-state wave function it is 
possible to obtain reasonably accurate 
results for the to properties of BEC, even beyond mean-field approach, without 
much of a computational effort.

\section{Conclusion}
In this paper we have studied the properties of Bose gas confined in both isotropic and 
axially symmetric potential going beyond GP or mean-field approximation by taking 
higher-order terms in the interatomic interaction energy. We have used the variational 
approach to solve the MGP equation for wide range of particle numbers. We have verified our
results using the generalized virial relation as well as by making analytic estimate of
the corrections introduced by higher-order terms. These higher-order 
terms lead to correction in the total energy, chemical potential and other physical 
properties of BEC. The magnitudes of these corrections are of the order of $1\%-2\%$
even for very large $N$. However, there is qualitative change in the density profile
of the condensate due to presence of the logarthimic term in the interaction energy.
We also critically examine the results in the literature and compare them with our numbers.
Here we emphasize that the variational method employed by us gives quite accurate results
with considerable computational ease.

It is well known that, corrections of this order are difficult to detect in the 
experimental \cite{stamper1}situation. However, these small changes are also reflected 
in the collective excitation frequencies of BEC and these quantities can be measured with 
greater accuracy.  Motivated by this we are now applying the variational approach and sum 
rule method of response theory of many-body systems \cite{bohigas,strinagiri} to calculate 
collective excitation frequencies. These results will be presented in our future 
publication.

\newpage
\begin{figure}[htb]
\begin{center}
\epsfig{figure=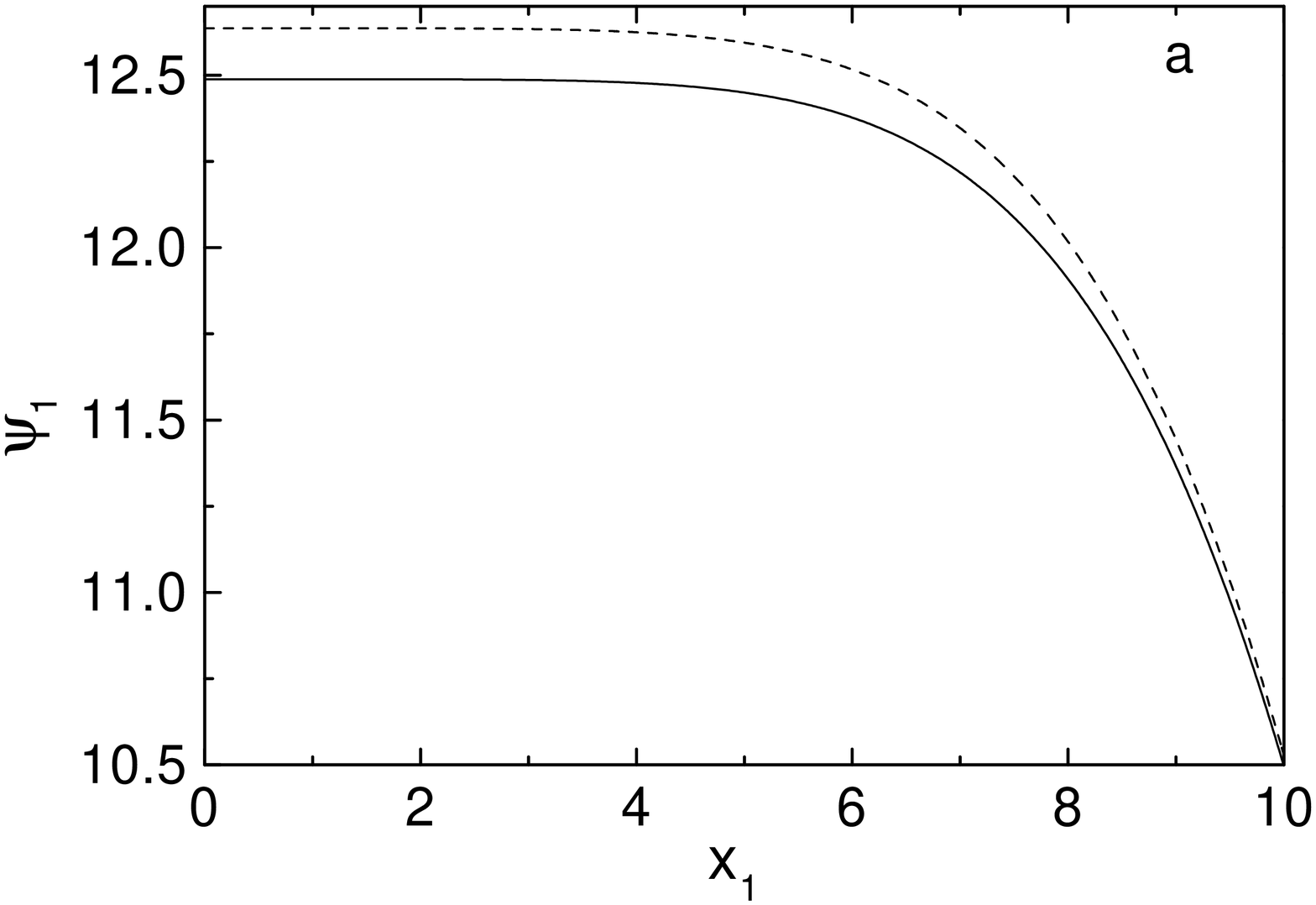,height=7.5cm,width=6.cm,angle=0}
\epsfig{figure=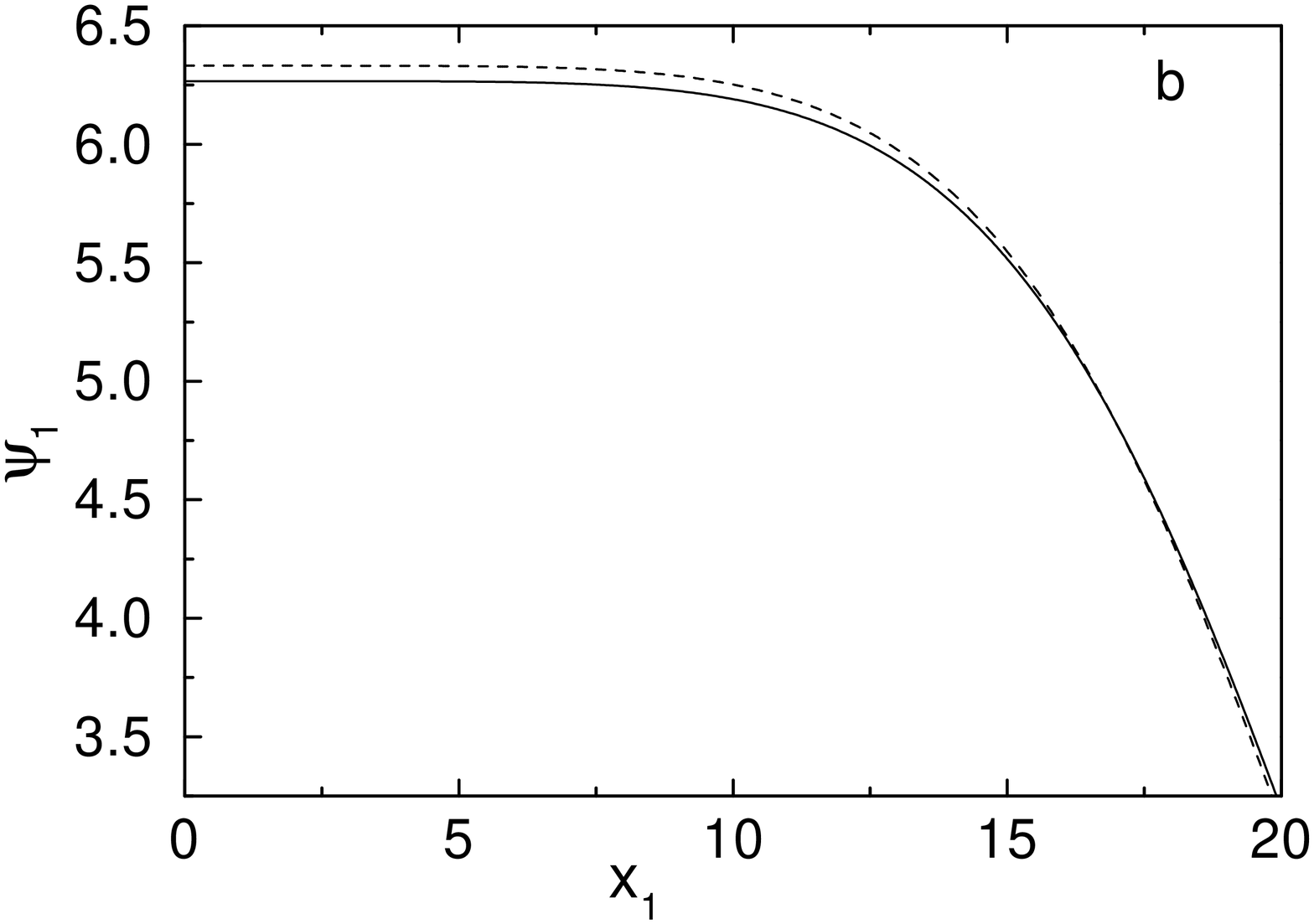,height=7.5cm,width=6.cm,angle=0}
\end{center}
\begin{center}
\epsfig{figure=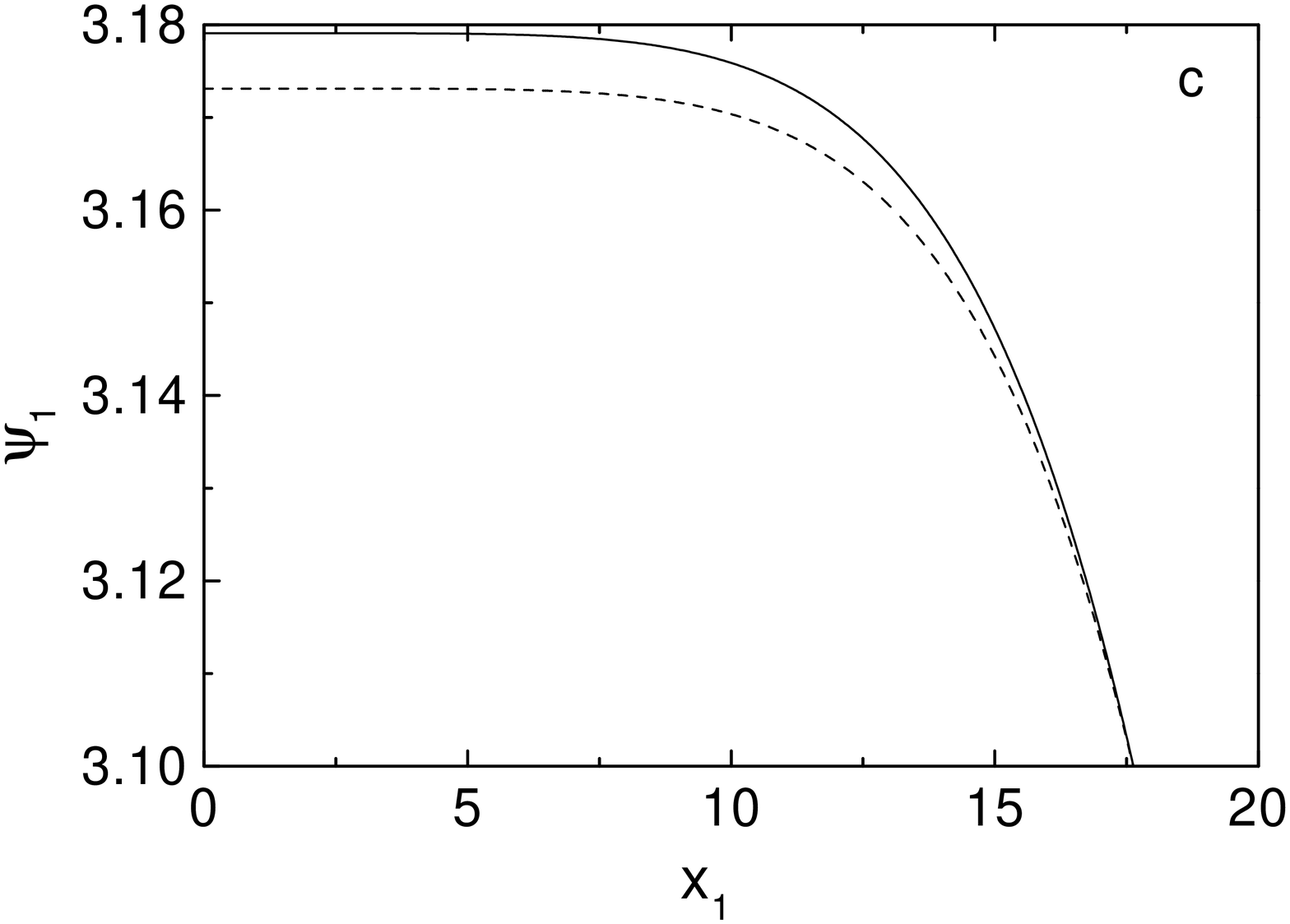,height=7.5cm,width=6.cm,angle=0}
\end{center}
\caption{ Plot of the wavefunction along transverse direction for different values of $N$: (a)
$N = 10^{7}$, (b) $N = 10^{8}$ and (c) $N = 10^{9}$. The solid and dashed curves 
correspond to solutions of GP and MGP equations respectively. The numbers along Y-axis are
multiplied with a scaling factor of 1000 } 

\end{figure} 
\clearpage
\pagebreak

\end{document}